\documentclass[a4paper,12pt]{article}
\usepackage{amsmath,amssymb,amsthm,mathrsfs,graphicx,float,colordvi,url,wrapfig,color,ulem}
\usepackage[all]{xy} 
\usepackage{tikz-cd} 

\makeatletter
\newcommand{\xleftrightarrow}[2][]{\ext@arrow 3359\leftrightarrowfill@{#1}{#2}} 
\newcommand{\xdashrightarrow}[2][]{\ext@arrow 0359\rightarrowfill@@{#1}{#2}}
\newcommand{\xdashleftarrow}[2][]{\ext@arrow 3095\leftarrowfill@@{#1}{#2}} 
\newcommand{\xdashleftrightarrow}[2][]{\ext@arrow 3359\leftrightarrowfill@@{#1}{#2}} 
\def\rightarrowfill@@{\arrowfill@@\relax\relbar\rightarrow}
\def\leftarrowfill@@{\arrowfill@@\leftarrow\relbar\relax}
\def\leftrightarrowfill@@{\arrowfill@@\leftarrow\relbar\rightarrow}
\def\arrowfill@@#1#2#3#4{
$\m@th\thickmuskip0mu\medmuskip\thickmuskip\thinmuskip\thickmuskip
\relax#4#1 
\xleaders\hbox{$#4#2$}\hfill
#3$% 
} 
\makeatother
\def\p{\partial}

\begin{document}
\renewcommand{\thefootnote}{\fnsymbol{footnote}} 
\begin{titlepage}

\vspace*{5mm}

\begin{center}
{\large 
\textbf{A 4D asymptotically flat rotating black hole solution
} 
\\ 
\vspace{1mm}
\textbf{including supertranslation correction}
}

\vspace*{8mm}

\normalsize
{\large Shingo Takeuchi}
%\footnote{shingotakeuchi@duytan.edu.vn}}

\vspace*{6mm} 

\textit{
\small Institute of Research and Development, Duy Tan University, Da Nang, Vietnam}\\
\vspace*{0.5 mm}
\textit{
\small Faculty of Environmental and Natural Sciences, Duy Tan University, Da Nang, Vietnam}\\

\end{center}

\vspace*{5mm}
\begin{abstract} 
In this study, 
beginning with the 4D asymptotically flat Kerr black hole solution 
in the Boyer-Lindquist (BL) coordinate system, 
then by using the technique of frame-dragging and some coordinate transformation 
to incorporate the linear-order supertranslation correction, 
we finally obtain a 4D asymptotically flat black hole solution with an arbitrary $a$ 
in the linear-order supertranslated BL coordinate system. 
We can confirm that this satisfies the Einstein equation. 
At the end, we calculate the angular momentum of the spacetime we obtained in this study 
including the linear-order correction. 
Since the supertranslation is a type of the general coordinate transformation, 
which is usually defined by linear-order, 
the corrections we incorporate in this study is sufficient, considering the fact that $a$ is arbitrary. 
This work is interesting as a generalization of the supertranslated spacetime to the rotating system.

\end{abstract}
\end{titlepage}

\newpage 
%******************* 
\allowdisplaybreaks 
%*******************

%=============================================
\section{Introduction} 
\label{sec:int} 
%============================================= 

BMS symmetry \cite{Bondi:1962px,Sachs:1962wk} is generated by the coordinate transformation  
defined in the vicinity of the infinite null region in the 4D asymptotically flat spacetime. 
In the BMS symmetry, symmetries which are spontaneously broken exist infinitely.  
It is interesting that, instead of the naive Poincar{\`{e}} symmetry, such symmetric structures
exist in the vicinity of infinite null region of the spacetime. 
In this paper, we refer to the BMS symmetry as the asymptotic symmetry.

It has been shown in~\cite{Barnich:2009se} that 
the asymptotic symmetry is given 
by the direct product of Virasolo symmetry 
on the conformal sphere in the vicinity of the infinite null region 
and the supertranslation along the retarded/advanced time direction at that conformal sphere.  
Thereafter, the central charges of the asymptotic symmetry have been analyzed~\cite{Barnich:2010eb,Barnich:2011mi}. 
On the other hand, the following two directions have been investigated: 
{\bf i)} the relation between the gravitational $S$-matrix and the gravitational soft-theorem~\cite{
Strominger:2013jfa,He:2014laa,Cachazo:2014fwa,Kapec:2014opa,Kapec:2016jld,He:2017fsb,Strominger:2013lka,He:2014cra,He:2019jjk}, 
and the holographic equivalence of the CFT with the 4D asymptotically flat spacetime~\cite{He:2015zea,Bagchi:2016bcd,
Pasterski:2016qvg,Cardona:2017keg,Pasterski:2017kqt,Pasterski:2017ylz}, and {\bf ii)}  
the resolution of the black hole information paradox, 
in which the missing information is supposed to be restored with the apparent deformation of the spacetime 
by the asymptotic symmetry~\cite{Hawking:2016msc,Hawking:2016sgy,Strominger:2017aeh}. 

Whether or not this asymptotic symmetry can be observed was discussed in~\cite{Sarkar:2021djs}. 
As a concrete observational methodology, the observation of the black hole shadow 
in the presence of supertranslation was proposed~\cite{Lin:2022ksb}.

The physical meanings of gravitational events and observations are changed 
for the coordinate transformation of the asymptotic symmetry 
(one of the reasons for this is that 
the time-axis at each point gets local transformations~\cite{Strominger:2017zoo,Compere:2018aar}), 
which is closely related to the deformation of spacetime 
which arises due to the propagation of the gravitational wave. 
%---
Therefore, now that the gravitational wave has been observed~\cite{LIGOScientific:2016aoc},  
the observation of the effect of the asymptotic symmetry 
to be read from the observation of the gravitational wave 
is gathering a lot of  attention of the observational gravity.
%--- 
One such phenomenon is the gravitational memory, 
which is an observable to measure the variation of distance between two objects for the propagation of the gravitational wave, 
which is closely related to the asymptotic symmetry~\cite{Strominger:2014pwa}. 

The gravitational memory in various systems has been studied, for example, 
spin memories~\cite{Pasterski:2015tva},
color memories~\cite{Pate:2017vwa}, 
and electromagnetic memories~\cite{Bieri:2013hqa,Susskind:2015hpa,Pasterski:2015zua}.

Astrophysical studies on the gravitational memory in the gravitational waves 
from merging binary black holes and in other backgrounds have been studied in~\cite{
Pshirkov:2009ak,vanHaasteren:2009fy,NSeto, Favata:2010zu,Pollney:2010hs,Wang:2014zls,Arzoumanian:2015cxr,Bieri:2015yia,
Boyle:2015nqa,Lasky:2016knh,McNeill:2017uvq, NANOGrav:2019vto, Hubner:2019sly,Mitman:2020bjf,Zhao:2021hmx} 
and~\cite{Madler:2017umy,Nichols:2017rqr,Madison:2017nhl,Talbot:2018sgr,Jenkins:2021kcj,Divakarla:2021xrd,Tahura:2021hbk,Ferko:2021bym}, respectively. 

Currently, the asymptotic symmetry has only been defined in the asymptotically flat non-rotating spacetime. 
Therefore, analyses in observational studies usually assume that for the background as well. 
However, gravitational sources are generally in rotation.  
Therefore, in this paper, we attempt to obtain an asymptotically flat rotating black hole solution with supertranslation corrections. 
\newline

The key methodology and basic outline of our study is 
to first use the frame-dragging technique~\cite{wikikerr} 
to transform the rotating black hole metric to a non-rotating one. 
Then, we incorporate the supertranslation corrections 
by using the coordinate transformation including the supertranslation corrections given in \cite{Compere:2016hzt}. 
%By denoting the non-rotating coordinate system obtained using the frame-dragging technique 
%with the original rotating coordinates,
Then, we revert it to the original rotating system.  
As a result, we finally obtain
a rotating black hole solution in the coordinate system with supertranslation corrections.

However, the black hole rotation parameter $a$ in the supertranslated rotating black hole solution above is at the linear-order. 
This is because $a$ is taken to the linear-order when the frame-dragging technique is used. 
However, in the process to obtain the above solution, we can obtain some coordinate transformation rules
to incorporate the supertranslation corrections into the rotating black hole solution. 
Therefore, applying that rule to the first 4D asymptotically flat Kerr black hole solution, 
we finally obtain a 4D asymptotically flat black hole solution 
with an arbitrary $a$ 
in the linear-order supertranslated Boyer-Lindquist (BL) coordinate system.
We confirm that this satisfies the Einstein equation. 
What we do in this study can be considered as a type of the generalization of the asymptotic symmetry to the rotating spacetime.

This generalization is not straightforward for the following reasons.  
%-----
The asymptotic symmetry is defined 
by the Killing vectors in the Bondi gauge 
in the vicinity of the infinite null region~\cite{Strominger:2017zoo}.  
However, metrices of rotating spacetimes could not be recast to that gauge. 
%-----
In addition, the effect of the spacetime rotation is very small 
in the vicinity of the infinite null region where the asymptotic symmetry is defined,   
and it is unclear how to extend the asymptotic symmetry to the middle region of spacetime 
(\cite{Liu:2022uox,Cai:2016idg} are examples of some studies 
on the asymptotic symmetry in the non-rotating near-horizon region).
%-----
Lastly, although this is common in the non-rotating case, 
one would usually have to work directly on the Killing vectors 
when performing analysis of the supertranslation, 
which makes analysis technically difficult.
It is likely that these are reasons 
why this generalization has not been performed until now.

In our study, we resolve these points
by using the frame-dragging technique, 
which enables us to treat the coordinate system of the rotating black hole as that of a non-rotating one, 
and the coordinate transformation to incorporate the supertranslation corrections. 

%---
We can consider that the linear-order supertranslation corrections 
we incorporate in this study are sufficient in principle, 
considering the fact that $a$ is arbitrary, 
where the corrections we treat in this study are 
$a$ and the supertranslation corrections.
This is because supertranslation is a type of the general coordinate transformation, 
which is usually defined by linear-order. 
We discuss this point more specifically in Sec.\ref{sbrevre}. 
\newline

Our paper is organized as follows:
%---
In Sec.\ref{nveoa}, beginning with the 4D rotating black hole spacetime 
in the BL coordinate system, 
we rewrite this into a Schwarzschild-type solution in an isotropic coordinate system 
by keeping $a$ up to the linear order using the frame-dragging technique. 
%---
In Sec.\ref{nvsdkq}, 
we incorporate the $\varepsilon^1$-order supertranslation correction in the coordinate system
by using the coordinate transformation. 
%---
In Sec.\ref{sbrevre}, we discuss that 
the linear-order supertranslation corrections we incorporate
in this study are sufficient. %, as noted above.  
%---
In Sec.\ref{subsec:rwscisc} and Sec.\ref{subsec:ormt}, 
we obtain a rotating black hole solution 
with the $a^1$-order black hole rotation 
in the $\varepsilon^1$-order supertranslated BL coordinate. 
%---
In Sec.\ref{elmgrgg}, by applying a coordinate transformation rule obtained in Sec.\ref{subsec:ormt}  
to the 4D rotating black hole spacetime in the BL coordinate system, 
we obtain a rotating black hole solution 
with an arbitrary $a$ in the $\varepsilon^1$-order supertranslated BL coordinate system.
%---
In Sec.\ref{nqcwei}, we calculate the angular momentum of the spacetime including the $\varepsilon^1$-order correction. 
%---
In Sec.\ref{Summary}, we summarize this work and mention our future work.  
%==========
Appendixes~\ref{cworfh}, \ref{g34q} and \ref{enlpu} are devoted 
to some details and calculation.
%-----
Appendix~\ref{cwenclwe} is devoted 
to review how to derive the coordinate transformation we use in this study 
to incorporate the linear order supertranslation.
%-----
Appendix~\ref{nowyt} is devoted 
to discuss some subtle points related to the frame-dragging technique.
%-----
In Appendix~\ref{rntre}, 
we show the results of the expansions of the metrices that we obtained in this study, 
which is helpful for seeing if it is feasible or not 
to incorporate $a$ and $\varepsilon$
in our future work with our metrices.  

%=============================================
\section{Asymptotically flat rotating black hole solution in the supertranslated coordinate}
\label{anrbh} 
%=============================================
%=============================================
\subsection{From the Kerr BH in the Boyer-Lindquist coordinate to Schwarzschild BH 
in the isotropic coordinate keeping $a$ up to linear order}
\label{nveoa} 
%=============================================

We begin with the 4D rotating black hole spacetime in the Boyer-Lindquist (BL) coordinate system,
\begin{eqnarray}\label{kcnh1} 
ds^2
\!\!\! &=& \!\!\!
-(1 - \frac{2m r}{r^2+ a^2 \cos^2 \theta})dt^2
+\frac{r^2+ a^2 \cos^2 \theta}{r^2-2m\,r+a^2}dr^2
+(r^2+ a^2 \cos^2 \theta) d\theta^2\nonumber\\*
&&
+( r^2+a^2+\frac{2m r a^2 \sin^2 \theta}{r^2+a^2\cos^2 \theta} ) \sin^2 \theta d\phi^2-\frac{4m r a \sin^2 \theta}{r^2+ a^2 \cos^2\theta}dt d \phi. 
\end{eqnarray}
We incorporate the transformation of the supertranslation in the coordinate system above up to the linear order.

For this purpose, let us remember that 
there is a method to incorporate the supertranslation 
in the isotropic coordinate system 
in \cite{Compere:2016hzt},
which is defined in the isotropic coordinate system 
and considered to be inapplicable to (\ref{kcnh1}) 
(for this, see Appendix.\ref{cworfh}). 
Therefore, in order to use it, 
we should once change the coordinate system of (\ref{kcnh1}) 
to the isotropic one.
Thus, we first rewrite (\ref{kcnh1}) in the Schwarzschild type coordinate system 
by keeping $a$ to the linear order as 
\begin{eqnarray}\label{aweq}
ds^2 = -( 1-{2m}/{r} ) dt^2 + (1-{2m}/{r})^{-1} dr^2 + r^2d\theta^2 + r^2 \sin^2 \theta d\varphi^2 +{\cal O}(a^2), 
\end{eqnarray}
where $a$ is implicitly contained in $d\varphi$ as 
\begin{eqnarray}\label{wvcuqq}
d\varphi = d\phi-\frac{2m \, a}{r^3}dt.
\end{eqnarray}
The transformation (\ref{wvcuqq}) is referred to as {\it the frame-dragging} \cite{wikikerr}. 
Then, it is straightforward to rewrite (\ref{aweq}) in the isotropic coordinate system:
\begin{eqnarray}\label{hoge}
ds^2 = - \frac{(1-{m}/{2\rho})^2}{(1+{m}/{2\rho})^2}dt^2 + ( 1+{m}/{2\rho} )^4 (d\rho^2+\rho^2(d\theta^2+\sin \theta^2 d\varphi^2))+{\cal O}(a^2). 
\end{eqnarray}
where $r=\rho (1+{m}/{2\rho} )^2$. 

Now we have transformed the coordinate system to the isotoropic one (\ref{hoge}) 
for the reason mentioned above. 
However, in Sec.\ref{elmgrgg}, 
we find a coordinate transformation rule to incorporate the supertranslation 
which is considered to be applicable to (\ref{kcnh1}) based on a crosscheck (\ref{rebih}). 
Applying this to (\ref{kcnh1}), 
we finally obtain a linear-order supertranslated rotating black hole solution 
with an arbitrary $a$ as in (\ref{klafh}).
Here, up to the linear-order is sufficient regarding the supertranslation corrections. 
We discuss this point in Sec.\ref{sbrevre}.
~\newline

Here, there is a subtle point, 
which is that 
if $d\varphi$ were given as in (\ref{wvcuqq}), 
$\varphi$ would be given as $\phi-{2m \, a t}/{r^3}$ by integrating it. 
Then, since this $\varphi$ is a function of $t$, $r$ and $\phi$, 
it is led to
$d\varphi
=\frac{\partial \varphi}{\partial t}dt+\frac{\partial \varphi}{\partial r}dr+d \phi 
=-{2m \, a}/{r^3}dt+{6m \, at}/{r^4}dr+d \phi$, 
which does not agree with the original $d\varphi$ in (\ref{wvcuqq}),  
and leads to the problem that we cannot use the quantity $\varphi$ as the coordinate.

Regarding this point, 
as long as we are merely denoting the quantity $d\phi-\frac{2m \, a}{r^3}dt$, 
which is the r.h.s. of (\ref{wvcuqq}), as $d\varphi$, 
there is no problems, 
as this is just a problem of notation; 
the problem is the points in which $\varphi$ appears. 

With this fact, if we look at equations in this work, 
we can see that the points where $\varphi$ appears are the coordinate transformation rules (\ref{ctki0}), 
which are considered 
regarding the part ``$d\rho^2 + \rho^2 (d\theta^2 + \sin^2 \theta d\varphi^2)$'' in (\ref{hoge}) 
as the three-dimensional space  (3D part) 
independent of the original 4D spacetime;  
namely, it is the space irrelevant of $r$. 
Here, $r$ and $\rho$ play roles of radial directions in  the original 4D spacetime and 3D space, respectively.

Certainly, $r$ and $\rho$ are linked to each other, as can be seen under (\ref{hoge}).
However, when we consider performing (\ref{ctki0}), the analysis is limited regarding the 3D part\footnote{
%====================
%==== FOOTNOTE =====
%====================
``limited regarding the 3D part'' means as author's intention that 
it is possible to perform analysis only with things in the 3D part, only in the 3D part, 
independently of the original 4D spacetime.
%====================
} 
, 
and in such a situation, we can consider $r$ and $\rho$ separately.  

Therefore, we can consider that $\varphi$ in (\ref{ctki0}) can be given as $\phi-{2m \, a t}/{r^3}$. 
However, $r$ is not the independent variable in that $\varphi$; 
namely the term $\frac{\partial \varphi}{\partial r}dr$ will not appear in $d\varphi$ in (\ref{ctki0}), 
and $d\varphi$ in (\ref{ctki0}) is consistent with (\ref{wvcuqq}). 
We discuss this point in more detail in Appendix \ref{nowyt}.
\newline

Lastly, we mention the terminology we use in this paper. 
Since supertranslation is a type of coordinate transformation, 
in this paper, we usually express the transformation of the supertranslation as 
{\it displacement of coordinates} or something similar. 

%=============================================
\subsection{Schwarzschild BH (\ref{hoge}) in the linear order supertranslated coordinate}
\label{nvsdkq} 
%============================================= 

Denoting the Cartesian coordinates of this 3D part, 
which is defined in Sec.\ref{nveoa}, 
as $(x,y,z)$, we can rewrite those as follows: 
\begin{subequations}\label{bdml} 
\begin{align}
%-----
\label{bdml1} 
&
\rho = (x^2+y^2+z^2)^{1/2}, 
\\*[0.5mm]
%----- 
\label{bdml2} 
&
d\rho^2 + \rho^2 (d\theta^2 + \sin^2 \theta d\varphi^2) = dx^2+dy^2+dz^2. 
\end{align}
\end{subequations}
We stress that these $(x, y, z)$ are not Cartesian coordinates of the original 4D spacetime, 
which can be seen from the fact that its radial coordinate is not $r$ but $\rho$.

Displaced $(t, x, y, z)$, 
which are occurred by the supertranslation, 
and which we denote as $(t_s, x_s, y_s, z_s)$, 
are given as follows:
\begin{subequations}\label{ctki0}
\begin{align}
x_s &= (\rho-C)\sin\theta \cos\varphi+\sin \varphi \csc \theta \,\partial_\varphi C-\cos\theta\cos\varphi \, \partial_\theta C, \\*
y_s &= (\rho-C)\sin\theta \sin\varphi- \cos \varphi \csc \theta \,\partial_\varphi C-\cos\theta\sin\varphi \, \partial_\theta C, \\*
z_s &= (\rho-C)\cos\theta+\cos\theta\cos\varphi \, \partial_\theta C, 
\end{align} 
\end{subequations}
where $t_s=t$, and $C(\theta, \varphi)$ is a function to characterize the supertranslation~\cite{Compere:2016hzt}, 
which we can take arbitrarily as long as it is some function with regard to only the angle direction.
We review how (\ref{ctki0}) can be obtained in Appendix~\ref{cwenclwe}. 

We take $C$ as follows:
\begin{eqnarray}\label{cdsvds} 
C = m\varepsilon\, Y_2^0 =m\varepsilon\, \sqrt{\frac{5}{16\pi}}(3\cos^2\theta-1). 
\end{eqnarray} 
We have three remarks on (\ref{cdsvds}):
\begin{itemize}
\item
$\varepsilon$ is an infinitesimal dimensionless parameter for our counting of the order of the supertranslation we incorporate. 

\item
$m$ is the same as that in (\ref{hoge}), 
which we include in $C$ so that  it can have the same dimension as $\rho$ (where $G/c^2=1$ in this paper).

\item
The reason why we consider $Y_2^0$ is that it is supposed to be the most dominant mode 
in the quasi-normal modes (ring down waves)
emitted during the process by which a soft-hairy black hole is formed \cite{Berti:2005ys}\footnote{
%=====================
%===== FOOTNOTE =====
%=====================
The author is thankful to Vitor Cardoso for this information.
%=====================
%=====================
%=====================
}. 
\end{itemize}

Using (\ref{ctki0}) and (\ref{cdsvds}), 
we can write $\rho$ and the 3D part in (\ref{hoge}) 
in the coordinate system displaced by the supertranslation up to the $\varepsilon^1$-order as
\begin{subequations}\label{bwenl} 
\begin{align}
%-----
\label{bwenl1} 
& \bullet \hspace{2mm}
\rho_s 
=
(x_s^2+y_s^2+z_s^2)^{1/2} 
=
\rho -\frac{\varepsilon}{8} \sqrt{\frac{5}{\pi }} m (3 \cos (2 \theta )+1)+{\cal O}(\varepsilon^2), \\* 
%----- 
\label{bwenl2} 
& \bullet \hspace{2mm}
d\rho_s^2 + \rho_s^2 (d\theta_s^2 + \sin^2 \varphi_s d\varphi_s^2) 
= dx_s^2+dy_s^2+dz_s^2
\nonumber \\* 
%----- 
&
\hspace{9.5mm}
=\frac{1}{2} \rho ^2 \left(2 d\theta^2+d\phi^2\right)+d\rho^2-\frac{1}{2} d\phi^2 \rho ^2 \cos (2 \theta )
\nonumber \\* 
%----- 
&
\hspace{14.0mm}
+\frac{1}{16} \sqrt{\frac{5}{\pi }} \varepsilon m \rho (4 \cos (2 \theta ) (9 d\theta ^2-d\varphi ^2) -4 d\theta ^2
+(7-3\cos (4 \theta ))d\varphi ^2), 
\end{align}
\end{subequations}
where $\rho_s$, $\theta_s$ and $\varphi_s$ mean the supertranslated $\rho$, $\theta$ and $\varphi$. 
%---
Here, although we have reached (\ref{bwenl}) from (\ref{bdml}) with the transformation rule 
by the Cartesian coordinates (\ref{ctki0}), 
it is possible even by (\ref{vndswi}).  
We use (\ref{ctki0}), since it is technically more clearer regarding the corrections we are treating. 
%---
With these, we can write (\ref{hoge}) in the $\varepsilon^1$-order supertranslated coordinate system as
\begin{eqnarray}\label{yjtjr}
ds^2 \!\!\! &=& \!\!\! 
-\frac{(1-{m}/{2\rho_s})^2}{(1+{m}/{2\rho_s})^2}dt^2 + ( 1+{m}/{2\rho_s} )^4 (dx_s^2+dy_s^2+dz_s^2) 
\nonumber\\*
&\equiv& \! 
g_{tt}dt^2+g_{\rho\rho}d\rho^2+g_{\theta\theta}d\theta^2+g_{\varphi\varphi}d\varphi^2, 
\end{eqnarray}
where
\begin{eqnarray} 
& \hspace{-3.0mm}  \bullet& \hspace{-1.5mm} 
g_{tt} 
= 
-\frac{(m-2 \rho )^2}{(m+2 \rho )^2}
-\sqrt{\frac{5}{\pi }} \varepsilon m^2\frac{ (m-2 \rho )}{(m+2 \rho )^3}(3 \cos (2 \theta )+1)+{\cal O}(\varepsilon^2),
\nonumber\\*
%======================== 
& \hspace{-3.0mm}  \bullet& \hspace{-1.5mm} 
g_{\rho\rho} 
= 
\frac{(m+2 \rho )^4}{16 \rho ^4}
+\sqrt{\frac{5}{\pi }} \varepsilon m^2\frac{ (m+2 \rho )^3}{32 \rho ^5} (3 \cos (2 \theta )+1)+{\cal O}(\varepsilon^2),
\nonumber\\*
%======================== 
& \hspace{-3.0mm}  \bullet& \hspace{-1.5mm} 
g_{\theta\theta}
= 
\frac{(m+2 \rho )^4}{16 \rho ^2}
+ 
\sqrt{\frac{5}{\pi }} \varepsilon m (m+2 \rho )^3 \, 
\frac{ 3 \cos (2 \theta ) (5 m+6 \rho )+m-2 \rho }{64 \rho ^3}+{\cal O}(\varepsilon^2),
\nonumber\\*
%========================
& \hspace{-3.0mm}  \bullet& \hspace{-1.5mm} 
g_{\varphi\varphi}
= 
\frac{ (m+2 \rho )^4\sin ^2 \theta}{16 \rho ^2}
+\sqrt{\frac{5}{\pi }} \varepsilon m \frac{ (m+2 \rho )^3}{64 \rho ^3} \sin ^2\theta 
(\cos (2 \theta ) (9 m+6 \rho )+7 m+10 \rho)+{\cal O}(\varepsilon^2).
\nonumber
\end{eqnarray}
~\newline

We denote the notations employed in this paper.
\begin{eqnarray} 
&&\textrm{\hspace{-2.5mm} 
$\cdot$\, 
$g_{\mu\nu}$ 
for metrices in the $\varepsilon^1$-order supertranslated isotropic coordinate system,} 
\nonumber\\*
&&\textrm{\hspace{-2.5mm} 
$\cdot$\, 
$j_{\mu\nu}$ 
for metrices in the $\varepsilon^1$-order supertranslated Schwarzschild coordinate system,} 
\nonumber\\*
&&\textrm{\hspace{-2.5mm} 
$\cdot$\, 
$J_{MN}$ 
for metrices in the $\varepsilon^1$-order supertranslated BL coordinate with linear order $a$,} 
\nonumber\\*
&&\textrm{\hspace{-2.5mm} 
$\cdot$\, 
$K_{MN}$ 
for metrices in the $\varepsilon^1$-order supertranslated BL coordinate with an arbitrary $a$,} 
\nonumber
\end{eqnarray}
where $\mu,\nu=t,\rho, \theta,\varphi$ ($\varphi$ is defined in (\ref{wvcuqq}))
and $M,N=t,\rho, \theta,\phi$. 

%============================================================ 
\subsection{The reason incorporating up to $\varepsilon^1$-order is sufficient}
\label{sbrevre} 
%============================================================

In this study, we incorporate the supertranslation correction up to the linear order ($\varepsilon^1$-order), 
which is sufficient for the following logic.\\ 
%-----
{\bf 1)}~$\zeta$ in (5.2.3) in \cite{Strominger:2017zoo} is the quantity up to the linear order.
This can be seen from the fact that 
Lie derivatives concerning the coordinate transformation of the supertranslation are defined with this $\zeta$ 
as in (5.2.2) in \cite{Strominger:2017zoo}. 
\vspace{1mm}\\
%-----
{\bf 2)}~Hence, 
$f$ in (5.2.3) in \cite{Strominger:2017zoo} is the quantity up to the linear order as well. 
Then, since ${\cal L}_f C=f$ as in (5.2.6) in \cite{Strominger:2017zoo}, 
$C$ is also the quantity up to the linear order. 
\vspace{1mm}\\*
%-----
{\bf 3)}~The $C$ in (5.2.6) in \cite{Strominger:2017zoo} is the $C$ in (\ref{ctki0}) and (\ref{cdsvds}) in this study.
Hence, the title in this subsection can be concluded. 
%-----

%============================================================ 
\subsection{From a $\varepsilon^1$-order supertranslated isotropic coordinate (\ref{yjtjr}) 
to a $\varepsilon^1$-order supertranslated Schwarzschild coordinate}
\label{subsec:rwscisc} 
%============================================================

We have obtained metrices for Schwarzschild black hole spacetime 
in the $\varepsilon^1$-order supertranslated isometric coordinate system as in (\ref{yjtjr}). 
In this subsection, we transform its coordinate system 
to the $\varepsilon^1$-order supertranslated Schwarzschild system and obtain metrices for that coordinate system. 
For this purpose, we rewrite this part in (\ref{yjtjr}), $g_{tt}dt^2+g_{\rho\rho}d\rho^2$, 
to the following form: 
\begin{eqnarray}\label{schcdst}
-( 1-{2 \mu(\rho)}/{r} ) dt^2 
+(1-{2 \mu(\rho)}/{r})^{-1}dr^2+ \cdots,
\end{eqnarray} 
where ``$\cdots$'' denotes terms irrelevant in the analysis for now, 
on which we comment in Appendix \ref{g34q}.
For this purpose, 
we obtain the relation between $r$ and $\rho$, 
and $\mu(\rho)$, by solving the following equations:
\begin{subequations}
\begin{align}
\label{ccisa1}
\bullet & \quad \!\! - ( 1-{2 \mu(\rho) }/{r} ) =\, g_{tt},\\*
\label{ccisa2} 
\bullet & \quad \!\! \frac{1}{1-{2 \mu(\rho) }/{r}} \Big( \frac{dr}{d\rho} \Big)^2 =\, g_{\rho\rho}.
\end{align}
\end{subequations}

First, we can obtain the $r$ satisfying (\ref{ccisa1}) to the $\varepsilon^1$-order as follows:
\begin{eqnarray}\label{ctfits} 
\!\!
r
\!\!\! &=& \!\!\!
\frac{\mu (\rho ) (m+2 \rho )^2}{4 m \rho } 
+\frac{\varepsilon }{32 \rho ^2}\sqrt{\frac{5}{\pi }} (3 \cos (2 \theta )+1) \mu (\rho ) (m^2-4 \rho ^2)
+O(\varepsilon ^2). 
\end{eqnarray}
Then, let us obtain the $\mu(\rho)$. 
For this purpose, plugging (\ref{ctfits}) into the $r$ in (\ref{ccisa2}), 
solve this regarding $\mu(\rho)$ order by order up to the $\varepsilon^1$-order. 
\vspace*{0.5mm}
As a result, we can obtain 
\vspace*{0.5mm}
$\displaystyle 
\mu(\rho) 
=
m
+{c_1 \rho \varepsilon}/{(m+2 \rho )^2}
+O(\varepsilon ^2)$, 
where we identified the integral constant at the $\varepsilon^0$-order as 
$m$, and $c_1$ is the integral constant at the $\varepsilon^1$-order, 
which should vanish, otherwise $\mu(\rho)$ is divergent at $\rho=0$ when $m=0$. 
Therefore, $\mu(\rho)$ is finally obtained as
\begin{eqnarray}\label{ldafyj}
\mu(\rho)=m.
\end{eqnarray}

However, what we want to do is rewrite from the supertranslated isotropic coordinate system
to the supertranslated Schwarzschild coordinate system. 
Therefore, it is more useful to obtain the relation between $\rho$ and $r$ in the form ``$\rho = \cdots$'' inverting (\ref{ctfits}).
For this purpose, we plug $m$ into the $\mu$ in (\ref{ctfits}) according to (\ref{ldafyj}),  
then expand it up to the $\varepsilon^1$-order, and solve it in terms of $\rho$.
As a result, we can obtain the following $\rho$:\footnote{
%====================
%=====FOOTNOTE=====
%====================
In fact, there are two roots as
$
\rho^{(\pm)} 
=
\frac{1}{2} 
( \pm \sqrt{r (r-2 m)}-m+r)
+\frac{1}{8} \sqrt{\frac{5}{\pi }} \varepsilon m (3 \cos (2 \theta )+1)
+O(\varepsilon ^2).$ 
We take $\rho^{(+)}$ as in (\ref{trrlcx1}) based on the behavior of the leading part in the $\varepsilon$-expansion at large $r$.  
}
\begin{eqnarray}
%====================
%====================
%====================
\label{trrlcx1}
\rho
=
\frac{1}{2} 
( \sqrt{r (r-2 m)}-m+r)
+\frac{1}{8} \sqrt{\frac{5}{\pi }} \varepsilon m (3 \cos (2 \theta )+1)
+O(\varepsilon ^2). 
\end{eqnarray}
~\newline

With
(\ref{trrlcx1}) 
and  
(\ref{ldafyj}), 
we can rewrite the metrices in the $\varepsilon^1$-order supertranslated isometric coordinate system 
in (\ref{yjtjr}) to those in the $\varepsilon^1$-order supertranslated Schwarzschild coordinate system.  
For this, let us formally denote this rewriting  as
\begin{eqnarray}\label{rwfits}
\!\!\!\!\!\! && \!\!\!\!
ds^2 = 
g_{tt}dt^2
+g_{\rho\rho}d\rho^2
+g_{\theta\theta}d\theta^2
+g_{\varphi\varphi}d\varphi^2
\nonumber\\* 
\!\!\!\!\!\! &\rightarrow& \!\!\!\!
-\Big(1-\frac{2\mu}{r} \Big) dt^2
+\Big(1-\frac{2\mu}{r} \Big)^{-1} dr^2
+\Big( g_{\theta\theta}+g_{\rho\rho}\Big(\frac{\partial \rho}{\partial \theta}\Big)^2 \Big)d\theta^2 
+2g_{\rho\rho}\frac{\partial \rho}{\partial r}\frac{\partial \rho}{\partial \theta}d\rho d\theta 
\nonumber\\*
&& \!\!\!\! + \, 
j_{\varphi\varphi}d\varphi^2. 
\end{eqnarray}
A few remarks in order:
\begin{itemize}
\item
$g_{\mu\nu}$ are given in (\ref{yjtjr}), 
\item
$\rho$ in the second line of (\ref{rwfits}) is replaced with (\ref{trrlcx1}),
\item
since $\rho$ in (\ref{trrlcx1}) is the function of $r$ and $\theta$,
$d\rho$ is given as $\frac{\partial \rho}{\partial r}dr+\frac{\partial \rho}{\partial \theta}d\theta$. 
$(\frac{\partial \rho}{\partial \theta}d\theta)^2$ is at the $\varepsilon ^2$-order, 
which is irrelevant in our analysis.
\end{itemize}
We can finally obtain an equation, as follows:
\begin{eqnarray}\label{rsso}
\textrm{(\ref{rwfits})} 
\,\,=\,\,
j_{tt} dt^2 
+j_{rr} dr^2
+ j_{\theta\theta} d\theta^2 
+ 2j_{r\theta} dr d\theta 
+ j_{\varphi\varphi}d\varphi^2, 
\end{eqnarray}
where
\begin{eqnarray}
%========== 
&& \bullet 
\hspace{2mm}
j_{tt} = -(1-\frac{2 m}{r})
+{\cal O}(\varepsilon^2),
\nonumber\\*
%========== 
&& \bullet 
\hspace{2mm}
j_{rr} = (1-\frac{2 m}{r})^{-1}
+{\cal O}(\varepsilon^2),
\nonumber\\*
%==========
&& \bullet 
\hspace{2mm}
j_{\theta\theta} = r^2+\frac{3 \sqrt{\frac{5}{\pi }} m \cos (2 \theta ) (\sqrt{r (r-2 m)}+r)^4}{2 (\sqrt{r (r-2 m)}-m+r)^3}\varepsilon 
+{\cal O}(\varepsilon^2), 
\nonumber\\*
%==========
&& \bullet 
\hspace{2mm} 
j_{r\theta} = -\frac{3 \sqrt{\frac{5}{\pi }} m \sin (2 \theta) (r - \sqrt{r (r-2m)})^4}{8 \sqrt{r (r-2 m)} (\sqrt{r (r-2 m)} - m + r )^3}\varepsilon
+{\cal O}(\varepsilon^2), 
\nonumber\\*
%==========
&& \bullet 
\hspace{2mm}
j_{\varphi \varphi} = r^2 \sin ^2\theta +\frac{3 \sqrt{\frac{5}{\pi }} m \sin ^2(2 \theta )(\sqrt{r (r-2 m)}+r)^4}{8 (\sqrt{r (r-2 m)}-m+r)^3} \varepsilon
+{\cal O}(\varepsilon^2). \nonumber
\end{eqnarray} 
We can check that these satisfy the Einstein equation up to the $\varepsilon^1$-order.

%============================================================ 
\subsection{A BH solution with $a^1$-order rotation in the linear order supertranslated BL coordinate}
\label{subsec:ormt} 
%============================================================

In the previous subsection, we obtained metrices for a Schwarzschild black hole solution 
in the $\varepsilon^1$-order supertranslated Schwarzschild coordinate system as in (\ref{rsso}).  
In this subsection, 
we extend this construction to 
a rotating black hole solution in the supertranslated BL coordinate system 
with the linear order $\varepsilon$ and $a$.  
For this purpose, 
we revert  $\varphi$ in (\ref{rsso}) denoted using the frame-dragging technique (\ref{wvcuqq}).   
Finally, we can obtain the following:
\begin{eqnarray}\label{kpyui} 
ds^2=J_{tt} dt^2+ J_{rr}dr^2 +J_{\theta\theta}d\theta^2+J_{\phi\phi} d\phi^2+ 2J_{r\theta} drd\theta,
\end{eqnarray}
where $J_{MN}$ are given using $j_{\mu\nu}$ and $\Theta$ as
\begin{eqnarray}
J_{MN} \!=\!
\left( 
\begin{array}{llll}
j_{tt} +j_{\varphi \varphi} \Theta^2
& 0 & 0 
& j_{\varphi \varphi} \Theta \\
\\
0 
& j_{rr} 
& j_{r\theta }
& 0 \\ 
\\
0 
& j_{r\theta }
& j_{\theta \theta } 
& 0 \\
\\
j_{\varphi \varphi } \Theta
& 0 & 0
& j_{\varphi \varphi} 
\end{array}
\right) +{\cal O}(a^2) +{\cal O}(\varepsilon^2). \nonumber
\end{eqnarray}
Here, $\Theta_r$ is defined in (\ref{wvcuqq}) as $d\varphi \equiv d\phi + \Theta dt$, and 
\begin{eqnarray}
%========== 
&& \bullet 
\hspace{2mm}
J_{tt} = -1+{2 m}/{r},
\nonumber\\*
%========== 
&& \bullet 
\hspace{2mm}
J_{t\varphi} = %-1+{2 m}/{r} 
-\frac{2m\sin ^2\theta }{r}a
-\frac{3 \sqrt{\frac{5}{\pi }} m^2 \sin ^2(2 \theta ) (\sqrt{r (r-2 m)}+r)^4}{4 r^3 (\sqrt{r (r-2m)}-m+r)^3}a\,\varepsilon,
\nonumber\\*
%========== 
&& \bullet 
\hspace{2mm}
J_{rr} =(1-2 m/r)^{-1},
\nonumber\\*
%==========
&& \bullet 
\hspace{2mm}
J_{r\theta} =-\frac{3 \sqrt{\frac{5}{\pi }} m \sin (2 \theta ) (\sqrt{r (r-2 m)}+r)^4}{8 \sqrt{r (r-2 m)}(\sqrt{r (r-2 m)}-m+r)^3} \varepsilon ,
\nonumber\\*
%==========
&& \bullet 
\hspace{2mm}
J_{\theta\theta} =r^2+\frac{3 \sqrt{\frac{5}{\pi }} m \cos (2 \theta ) (\sqrt{r (r-2 m)}+r)^4}{2 (\sqrt{r (r-2m)}-m+r)^3}\varepsilon ,
\nonumber\\*
%==========
&& \bullet 
\hspace{2mm}
J_{\varphi_s \varphi _s} = r^2 \sin ^2\theta +\frac{3 \sqrt{\frac{5}{\pi }} m \sin ^2(2 \theta ) (\sqrt{r (r-2 m)}+r)^4}{8 (\sqrt{r (r-2m)}-m+r)^3} \varepsilon. 
\nonumber
\end{eqnarray} 
We can check that these satisfy the Einstein equation up to the $\varepsilon^1$- and $a^1$-orders.

%=============================================
\subsection{An asymptotically flat BH solution with an arbitrary $a$ in an $\varepsilon^1$-order supertranslated BL coordinate}
\label{elmgrgg} 
%=============================================

We have obtained (\ref{rsso}) from (\ref{aweq}) via the isotropic coordinate (\ref{hoge}).  
In this subsection, we first obtain the coordinate transformation rule to directly transform (\ref{aweq}) to (\ref{rsso}). 
Then, applying it to (\ref{kcnh1}), 
we obtain an asymptotically flat black hole solution with an arbitrary black hole rotation $a$
in the $\varepsilon^1$-order supertranslated Schwarzschild coordinate system. 

First of all, let us displace $\theta$ and $\varphi$ 
with the $\varepsilon^1$-order supertranslation, 
which we denote as $\theta_s$ and $\varphi_s$, respectively. 
For this purpose, we use the stereographic map: $z_s=e^{i\varphi_s}\cot \theta_s/2$. 
The $z_s$ means the displaced $z$ by the $\varepsilon^1$-order supertranslation, which is given in terms of $z$ as 
\begin{eqnarray}\label{lfjsq}
z_s = 
\frac{(z \bar{z}-1) (\rho -C)+(z \bar{z}+1) (\rho_s-\bar{z} \partial_{\bar{z}} C-z \partial_z C)}
{2 \bar{z} (\rho -C)+(z \bar{z}+1)(\bar{z}^2 \partial_{\bar{z}} C-\partial_z C)},
\end{eqnarray} 
where the one above is given in (\ref{vndswi}), 
and $C$ and $\rho_s$ are given in (\ref{cdsvds}) and (\ref{bwenl2}). 
Then $\theta_s$ and $\varphi_s$ can be written in terms of $z_s$ as 
\begin{eqnarray}\label{ldsrtr}
\theta_s=2 \cot ^{-1}|z_s|, \quad \varphi_s=\frac{i}{2} \ln \bar{z_s}z_s^{-1}. 
\end{eqnarray} 

We plug $\rho_s$ in (\ref{bwenl2}) and $C$ in (\ref{cdsvds}) into the $z_s$ in (\ref{lfjsq}).
Here, $\rho_s$ and $C$ are given in terms of $z$ 
by rewriting $\theta$ and $\varphi$ in those in terms of $z$. 
Then, $\theta_s$ and $\varphi_s$ can be obtained as follows: 
\begin{subequations}
\begin{align}
\label{lhrrvb1}
\theta_s =& \,\, 
\theta +\frac{3m}{2 (r+\sqrt{(r-2m) r}-m)} \sqrt{\frac{5}{\pi }} \varepsilon \sin (2 \theta) +{\cal O}(\varepsilon^2)
\equiv 
\theta +\delta\theta, \\*
\label{lhrrvb2}
\varphi_s =& \,\, 
\varphi +{\cal O}(\varepsilon^2). 
\end{align}
\end{subequations}
By comparing (\ref{wvcuqq}) with (\ref{bwenl1}) and (\ref{lhrrvb2}), 
we can identify $\phi_s$ as
\begin{eqnarray}\label{lhrrvb3}
\phi_s=\phi +{\cal O}(\varepsilon^2). 
\end{eqnarray} 

Now, we are ready to directly obtain $J_{\mu\nu}$ from (\ref{aweq}) by performing the following transformation: 
\begin{eqnarray}\label{sklbwv}
(r, \theta,\varphi) \to (r_s, \theta_s,\varphi_s),
\end{eqnarray} 
where $r_s$ can be obtained from the relation given under (\ref{hoge}) with (\ref{bwenl1}), 
and $\theta_s$ and $\varphi_s$ are obtained as in (\ref{lhrrvb1}) and (\ref{lhrrvb2}), respectively. 
We can confirm that (\ref{aweq}) can be reduced to (\ref{rsso}) by applying (\ref{sklbwv}). Namely,
\begin{subequations}\label{rebih}
\begin{align}
\label{rebih1}
\!\!\!
\bullet \quad \!\!\!     
\textrm{(\ref{aweq})}                            
&  
\xrightarrow[\textrm{isometric coordinate}]{} 
\textrm{(\ref{hoge}):~Schwarzschild in isometric coord.} 
\nonumber\\*              
& 
\xrightarrow[\textrm{supertranslation (\ref{ctki0})}]{}      
\textrm{(\ref{yjtjr}):~supertranslated Schwarzschild in isometric coord.}
\nonumber\\*              
& 
\xrightarrow[\textrm{Schwarzschild form}]{} 
\textrm{(\ref{rsso}):~supertranslated Schwarzschild in Schwarzschild form,}
\\[1.5mm]
%-----
\label{rebih2}
\!\!\!
\bullet \quad \!\!\!     
\textrm{(\ref{aweq})}                                          
& 
\xrightarrow[\,\,\,\,\,\text{supertranslation (\ref{sklbwv})}\,\,\,\,\,]{}      
\textrm{(\ref{rsso}):~supertranslated Schwarzschild in Schwarzschild form.}
\end{align}
\end{subequations}
In (\ref{rebih1}), the isometric coordinate is gone through to reach (\ref{rsso}), 
while  in (\ref{rebih2}), (\ref{rsso}) is directly reached. 
Both of these can agree with each other, and actually do agree.

By applying (\ref{sklbwv}) to (\ref{kcnh1}), 
we can obtain an asymptotically flat black hole solution with an arbitrary $a$  
in the $\varepsilon^1$-order supertranslated BL coordinate system as 
\begin{eqnarray}\label{klafh} 
ds^2 = K_{MN} dx^M dx^N,
\end{eqnarray}
where
\begin{eqnarray} 
%========== 
&& \bullet 
\hspace{2mm}
K_{tt} = 
-1+\frac{2 m r}{a^2 \cos ^2\theta +r^2}
+\frac{3 \sqrt{\frac{5}{\pi }} a^2m^2 r \sin ^2(2 \theta )}{(r+\sqrt{(r-2m) r}-m) (a^2 \cos ^2\theta +r^2)^2} \varepsilon+{\cal O}(\varepsilon^2),
\nonumber\\*
%========== 
&& \bullet 
\hspace{2mm}
K_{t\phi} = 
-\frac{2 a m r \sin ^2\theta }{a^2 \cos ^2\theta +r^2}
-\frac{3 \sqrt{\frac{5}{\pi }} a m^2 r (a^2+r^2) \sin ^2(2 \theta )}{(r+\sqrt{(r-2m) r}-m) (a^2 \cos ^2\theta +r^2)^2}\varepsilon+{\cal O}(\varepsilon^2),
\nonumber\\*
%========== 
&& \bullet 
\hspace{2mm}
K_{rr} = \frac{a^2 \cos ^2\theta +r^2}{a^2-2 m r+r^2}
-\frac{3 \sqrt{\frac{5}{\pi }} a^2 m\sin ^2(2 \theta )}{2 (r+\sqrt{(r-2) r}-m) (a^2+r (r-2 m))} \varepsilon +{\cal O}(\varepsilon^2),
\nonumber\\*
%==========
&& \bullet 
\hspace{2mm}
K_{r\theta} = -\frac{3 \sqrt{\frac{5}{\pi }} m \sin (2 \theta ) (a^2 \cos ^2\theta +r^2)}{2 \sqrt{(r-2m) r} (r+\sqrt{(r-2m) r}-m)}\varepsilon+{\cal O}(\varepsilon^2),
\nonumber\\*
%==========
&& \bullet 
\hspace{2mm}
K_{\theta\theta} = 
a^2 \cos ^2\theta +r^2
+\frac{3 \sqrt{\frac{5}{\pi }}m (a^2 (3 \cos (4 \theta )+1)+4 (a^2+2 r^2) \cos (2 \theta ))}{4 (r+\sqrt{(r-2m) r}-m)} \varepsilon+{\cal O}(\varepsilon^2), 
\nonumber\\*
%==========
&& \bullet 
\hspace{2mm}
K_{\phi \phi} = 
\sin ^2\theta (\frac{2 a^2 m r \sin ^2\theta }{a^2 \cos ^2\theta +r^2}+a^2+r^2)
\nonumber\\*
&&\hspace{17mm}
+\frac{3 \sqrt{\frac{5}{\pi }} m \sin ^2(2 \theta )}{16 (r+\sqrt{(r-2m) r}-m) (a^2 \cos ^2\theta +r^2)^2}
\big\{(3 a^6+a^4 r (10 m+11 r)
\nonumber\\*
&&\hspace{17mm}
+16 a^2 r^3 (m+r)
+a^2 (a^2+r (r-2 m)) (a^2 \cos (4 \theta )+4 (a^2+2 r^2) \cos (2\theta ))
\nonumber\\*
&&\hspace{17mm}
+8 r^6\big\}\varepsilon+{\cal O}(\varepsilon^2).
\nonumber
\end{eqnarray} 
We can check that $K_{MN}$ above satisfies the Einstein equation 
up to the $\varepsilon^1$-order for an arbitrary $a$, and agrees with the slowly rotating metrices (\ref{kpyui}) 
if terms higher than the $a^1$-order are truncated in those. 
Further, $K_{MN}$ above agrees with the metrices of the supertranslated Schwarzschild black hole spacetime given in (45) in \cite{Takeuchi:2021ibg} at $a=0$. 

%=============================================
\section{The relation between spacetime and angular momentum}
\label{nqcwei}
%============================================= 

In this section we check the angular momentum of the spacetime we have obtained (\ref{klafh}) 
including the $\varepsilon^1$-order correction. 

For this purpose, we first consider some small gravitational perturbations 
from a stationary axisymmetric rotating object, in general, as
\begin{align}\label{celwr}
\eta_{\mu\nu}+h_{\mu\nu},
\end{align}
where $\eta_{\mu\nu}$ are the metrices for the flat Minkowski spacetime and $h_{\mu\nu}(t,r,\theta)$ are the perturbations. 
Note that $h_{\mu\nu}$ is independent of $\phi$, 
reflecting our assumption that the spacetime is axisymmetric toward $\phi$. 

Now, let us consider the Einstein equation, $R_{\mu\nu}-1/2 g_{\mu\nu} R = 8\pi T_{\mu\nu}$, in the spherical coordinates 
$(t,r,\theta,\phi)$ (therefore, $\eta_{\mu\nu} $ is given as ${\rm diag}(-1,1,r^2,r^2\sin^2 \theta)$). 
Then we can obtain the following equation for the $t\phi$-component:
\begin{align}\label{xqty}
\partial_r^2 h_{t\phi}+\frac{\sin \theta}{r^2}\partial_\theta (
\frac
{\partial_\theta h_{t\phi}}
{\sin \theta}
)
=-16\pi T_{t\phi}.
\end{align}
This equation can be rewritten as
\begin{align}\label{dcjic}
\triangle(\frac{\cos \phi}{\sin \theta}\frac{h_{t\phi}}{r})=
-16\pi \frac{\cos \phi}{r \sin \theta}T_{t\phi},
\end{align}
where $\triangle$ means the Laplacian on flat space. 
Therefore, we can get\footnote{
%=======================
%====== FOOTNOTE ======
%=======================
The Green function toward $\triangle$ is 
$-\frac{1}{4\pi}\frac{1}{|\vec{x}-\vec{x'}|}$, where $\triangle\frac{1}{|\vec{x}-\vec{x'}|}=-4\pi\delta^3(\vec{x}-\vec{x'})$.
%=======================
}
\begin{align}\label{sbdva}
\frac{\cos \phi}{\sin \theta}\frac{h_{t\phi}}{r}
=4\int_{\Omega} d^3\vec{x'} \frac{1}{|\vec{x}-\vec{x'}|} \frac{\cos \phi'}{r'\sin \theta'} T_{t\phi}(\vec{x'}),
\end{align}
where $\Omega$ denotes the region occupied by the rotating object, 
and $r=|\vec{x}|$, $r'=|\vec{x'}|$. 
If the stationary axisymmetric rotating object is very small compared with the $r$, 
(\ref{sbdva}) can be written as\footnote{
%=======================
%====== FOOTNOTE ======
%=======================
We note the derivation of (\ref{jiort}) from (\ref{sbdva}) in Appendix \ref{enlpu}.
%=======================
}
\begin{align}\label{jiort}
{\rm (\ref{sbdva})}=
\frac
{2\sin \theta \cos \phi}
{r^2}
\int_\Omega d^3\vec{x'} (1+\frac{3r'}{r}\cos \theta\cos \theta')T_{t\phi}(\vec{x'})
+{\cal O}(r^{-4}).
\end{align}
In general, the angular momentum for an astronomical body can be written as follows:
\begin{align}\label{ejhoer}
J=-c^{-1}\int_\Omega d^3\vec{x} \, T_{t\phi}.
\end{align}
Using this, we can obtain $h_{t\phi}$ as
\begin{align}\label{rskfda}
h_{t\phi} = - \frac{2\sin^2 \theta}{r} J' +{\cal O}(r^{-3}), \quad
J'=J-\frac{3\cos \theta}{r}\int_\Omega d^3\vec{x'} \, r' \,\cos \theta' \, T_{t\phi}(\vec{x'}). 
\end{align}
Let us stress that 
the $h_{t\phi}$ above is the general gravitational field 
sourced by a stationary axisymmetric rotating object,
at the far region. 

Now, we can compare (\ref{rskfda}) with our metrices (\ref{klafh}). 
Then, since the $t\phi$-component of our metrices (\ref{klafh}) can be expanded at infinite $r$ as
\begin{align}\label{vjare}
K_{t\phi}
=
-\frac{2 a m \sin ^2\theta }{r}
-6 \sqrt{\frac{5}{\pi}} a m \sin ^2\theta \cos ^2\theta \frac{\varepsilon }{r^2}
+{\cal O}(r^{-3}), 
\end{align}
(\ref{vjare}) can agree to (\ref{rskfda}) with the following identification: 
\begin{align}\label{tfhe}
a=\frac{J'}{m}-\frac{3 \sqrt{\frac{5}{\pi }} J \cos ^2\theta }{m} \frac{\varepsilon}{r}
+{\cal O}(r^{-2}). 
\end{align}
where $J'$ is given in (\ref{rskfda}).
In conclusion, the $a$ in our metrices (\ref{klafh}) can be related to the angular momentum of the spacetime $J$ 
with the $\varepsilon^1$-order correction.

%=============================================
\section{Summary and future work} 
\label{Summary}
%============================================= 

In this study, we have obtained the metrices of an asymptotically flat rotating black hole spacetime solution 
with an arbitrary $a$ (black hole rotation)
in the linear-order ($\varepsilon^1$-order) supertranslated BL coordinate system as in (\ref{klafh}). 
We can confirm that (\ref{klafh}) satisfies the Einstein equation to the $\varepsilon^1$-order for an arbitrary $a$.
Lastly, we have obtained the angular momentum of the spacetime including the $\varepsilon^1$-order correction. 

As we explain in Sec.\ref{sbrevre}, 
since the supertranslation is a type of the general coordinate transformation, 
which is defined to the linear order in general, 
it is sufficient if we incorporated supertranslation corrections to the linear-order. 
Therefore, the corrections we have incorporated in this study are sufficient in principle,  
considering the fact that $a$ is arbitrary, 
where the corrections we have treated in this study are 
$a$ and $\varepsilon$.

Currently, the study of the asymptotic symmetry is limited in the asymptotically flat non-rotating spacetimes 
(see Sec.\ref{sec:int} regarding the reasoning  for this consideration). 
Therefore, the observational theory 
concerning the effect of the asymptotic symmetry in a gravitational wave, 
such as gravitational memory, also always assumes that for its background spacetime. 
However, since spacetimes of gravitational sources is rotating in general, 
extending the asymptotic symmetry to the asymptotically flat rotating spacetime 
is an important issue.
This work is interesting in this sense, 
because it has succeeded in 
the construction of a linear order supertranslated asymptotically flat rotating black hole solution 
with an arbitrary rotation parameter $a$.

As to our future work, 
we plan the gravitational memory 
in rotational spacetimes using the metrices we have obtained in this study. 
Actually, if we expanded our metrices, as shown in Appendix \ref{rntre}, 
it would be possible to perform the analysis 
concerning the gravitational memory in the Kerr spacetime with supertranslation corrections  
using our metrices. 
This will be considered for future research.

\appendix
%=============================================
\section{Inapplicability of the coordinate transformation to incorporate supertranslation} 
\label{cworfh}
%============================================= 

The author performed some analysis, 
which is that, 
vanishing $a$ in (\ref{kcnh1}), 
then switching it to the Kerr-Schild (KS) form, 
incorporate the supertranslation into that. 
Then backing it to the BL coordinate system, 
compare it with (\ref{rsso}). 
Then, the author found that these cannot agree each other. Namely,
\begin{subequations}\label{vthshorb}
\begin{align}
\label{vthsh1}
\!\!\!
\bullet \quad \!\!
\textrm{(\ref{kcnh1})}                            
& 
\xrightarrow[\textrm{$a \to 0$ and KS form}]{} 
\textrm{Schwarzschild in KS form} 
\nonumber\\*              
& 
\xrightarrow[\textrm{supertranslation}]{}      
\textrm{supertranslated Schwarzschild in KS form}
\nonumber\\*          
& 
\xrightarrow[\textrm{back to BL form}]{} 
\textrm{supertranslated Schwarzschild in Schwarzschild form,} \\*[1.5mm]
%==========
\label{vthsh2}
\!\!\!
\bullet \quad \!\!    
\textrm{(\ref{kcnh1})}                            
&  
\xrightarrow[a \,\to\, 0]{} 
\textrm{(\ref{aweq}):~Schwarzschild in Schwarzschild form,} 
\nonumber\\*              
& 
\xrightarrow[\textrm{(\ref{rebih1})}]{}      
\textrm{(\ref{rsso}):~supertranslated Schwarzschild in Schwarzschild form.}
\end{align}
\end{subequations}
In (\ref{vthsh1}) the supertranslation is incorprated in the KS form, 
while in (\ref{vthsh2}), in the isometric coordinate, 
where $a$ is taken to be zero to make the problem simple.
(\ref{vthsh1}) and (\ref{vthsh2}) should agree each other, however they do not.

Of course, the author checked technical and essential mistakes 
in the computations\footnote{
%==============================
%========== FOOTNOTE ==========
%==============================
For the coordinate transformation 
from the KS to BL forms,
the author mainly referred to the following 3: 
{\bf (ia)} (7.3) in \cite{Teukolsky:2014vca}, 
{\bf (ib)} (6) in \cite{Arkani-Hamed:2019ymq}, 
and {\bf (ic)} (34)-(36) in \cite{Visser:2007fj}. 
As a result, it was found that 
each and every one is slightly different from each other.

Next, the KS form is given in the following 2: 
{\bf (iia)} (1.2) in \cite{Gibbons:2004uw}, 
and {\bf (iib)} (32) in \cite{Visser:2007fj}. 
However, these are also slightly different from each other.

Finally, the author confirmed that
the combination of {\bf (ic)} and {\bf (iia)} can agree with the BL form, (3) in \cite{Visser:2007fj}. 
Therefore, basically, the analysis in the text body is done with these.
%==============================
%==============================
}, 
but none were found. 
Therefore, it was concluded that 
the method
in \cite{Compere:2016hzt} is applicable 
only in the isotropic coordinate system.

There is some possibility that
(\ref{vthsh1}) is right and (\ref{vthsh2}) is wrong, 
or neither (\ref{vthsh1}) nor (\ref{vthsh2}) are right.
As for this concern, in this study, 
(\ref{vthsh2}) is supposed to be right. 
Here, both (\ref{vthsh1}) and (\ref{vthsh2}) are satisfying the Einstein equation.  

%=============================================
\section{A review of the derivation (\ref{ctki0})} 
\label{cwenclwe}
%============================================= 

This Appendix reviews how to derive the coordinate transformation (\ref{ctki0}) 
given in \cite{Compere:2016hzt} 
when a 3D flat space is given.
\newline

The squared line element before the supertranslations in the flat spacetime in the retarded coordinate ($r$, $u$, $z$, $\bar{z}$) 
can be written as follows:
\begin{eqnarray}\label{cwemop}
ds^2=-du^2-2dudr+\frac{4r^2}{(1+z\bar{z})^2}dzd\bar{z},
\end{eqnarray}
which can be reduced to the polar coordinate $(t,r,\theta,\phi)$ by $u=t-r$ and $z=e^{i\phi}\cot \theta/2$. 
(\ref{cwemop}) can be obtained from
\begin{eqnarray}\label{eroip}
ds^2 = -2 du_c dr_c + 2 r_c^2 dz_c d\bar z_c
\end{eqnarray}
by the coordinate transformations:
\begin{eqnarray}\label{ctos}
r_c = \frac{\sqrt{2}r}{1+z\bar{z}}+\frac{1}{\sqrt{2}}u,\quad
u_c = \frac{1+z\bar{z}}{\sqrt{2}}u-\frac{z\bar{z}}{2 r_c}u^2, \quad 
z_c = z-\frac{z}{\sqrt{2}r_c}u. 
\end{eqnarray}

Then, a supertranslated (\ref{eroip}) is obtained by the following $r'_c$, $u'_c$ and $z'_c$:
\begin{subequations}\label{wgwep}
\begin{align}
r'_c=& \frac{\p_z \p_{\bar z} W}{\p_z G\p_{\bar{z}}G} +\sqrt{\frac{r^2}{(\p_u W)^2}
+\frac{(\p_z^2G\p_zW-\p_z G\p_z^2 W)(\p_{\bar{z}}^2\bar{G}\p_{\bar{z}}W-\p_{\bar{z}}\bar{G} \p_{\bar z}^2 W)}{(\p_zG)^3(\p_{\bar{z}}\bar{G})^3}},\\*
u'_c=& W - \frac{1}{r_c} \frac{\p_z W \p_{\bar z} W }{\p_zG\p_{\bar{z}}\bar{G}},\label{ctoBMS}\\*
z'_c=& G- \frac{1}{r_c}\frac{\p_{\bar z} W}{\p_{\bar{z}}\bar{G}},\quad 
\bar{z}_c' = \bar{G}- \frac{1}{r_c}\frac{\p_{z} W}{ \p_z G}.
\end{align}
\end{subequations} 
Here, $W(u, z, \bar{z})$ characterizes the Weyl rescalings and supertranslations, 
and $G(z)$ characterizes the Lorentz transformations. 

Let us take the $G$ and $W$ as follows:
\begin{eqnarray}\label{WW}
G= z,\quad
W= \sqrt{ \gamma_{z\bar{z}} } (u + C),
\end{eqnarray}
where $C(z, \bar{z})$ means the effect of the supertranslation, which we can take arbitrarily. 
(\ref{wgwep}) can be reduced to (\ref{ctos}) at $C=0$.
With (\ref{wgwep}) and  (\ref{WW}), we can obtain the supertranslated (\ref{eroip}) as
\begin{eqnarray}\label{eneip}
ds^2 \!\! &=& \!\! -du^2-2dudF+ H dz^Adz^B, \\
F \!\! &=& \!\! \sqrt{r^2+U} + \frac{1}{2}(D^2+2)C, \quad
H = (r^2 + 2U)\gamma_{AB} + \sqrt{r^2+U} C_{AB}, \nonumber 
\end{eqnarray}
where 
\begin{eqnarray}
z^A \!\! &=& \!\! (z,\bar{z}), \quad
C_{AB} = -(2D_AD_B-\gamma_{AB} D^2)C, \quad 
U=\frac{1}{8} C_{AB}C^{AB}, \quad 
\gamma_{AB}=\gamma_{z\bar{z}}^{-2}
\left(
\begin{array}{cc}
0 & 1 \\ 1 & 0 
\end{array}
\right) \nonumber
\end{eqnarray}
and $D_A$ is the covariant derivative. 
Note $D^2C=\gamma^{AB}D_AD_B C =\gamma^{AB}D_A \partial_B C$.
%---
With the following coordinate transformations:
\begin{eqnarray}
t = u + \rho,\quad
\rho = \sqrt{r^2+U} +E,\quad E = \frac{1}{2}D^2C +C.
\end{eqnarray}
(\ref{eneip}) can be further rewritten into the following equation: 
\begin{eqnarray}\label{metbwi}
{\rm (\ref{eneip})}=
-dt^2 +d\rho^2 + ( ((\rho- E)^2 + U)\gamma_{AB} + (\rho - E) C_{AB} )dz^Adz^B.
\end{eqnarray}
Here, we can check that (\ref{metbwi}) can be reduced to the Bondi gauge: 
\begin{align}
ds^2 = e^{2 \beta} \frac{V}{r}du^2 - 2 e^{2\beta} du dr + g_{AB} (dz^A - U^A du)(dz^B - U^B du),
\end{align}
where $\partial_r (r^{-2}\det (g_{AB}))=0$, $g_{rr}=g_{rz}=g_{r\bar{z}}=0$ and 
\begin{align}
g_{AB}
=r^2 \gamma_{AB}
\gamma_{z\bar{z}}^{-2}
\left(
\begin{array}{cc}
0 & 1 \\ 1 & 0 
\end{array}
\right)
+{\cal O}(r). 
\end{align}
~\newline

For the vanishing $C$, 
(\ref{metbwi})  can be reduced to the flat spacetime
\begin{eqnarray}\label{mrhieu}
\label{mrhieu1}
ds^2
=-dt^2 +dr^2
+ r^2\gamma_{AB} dz^Adz^B,
\end{eqnarray}
We can check that (\ref{mrhieu1}) can be mapped to (\ref{metbwi}) by the following replacement:
\begin{eqnarray}\label{vndswi}
(t, r, z) \to (t_s, \rho_s, z_s)
\end{eqnarray}
where
\begin{subequations}\label{vndswi}
\begin{align}
\label{vndswi1}
t_s \,=& \,\, t, 
\\*
\label{vndswi2}
\rho_s \,= & \,\, \sqrt{(\rho - C )^2 + D_A C D^A C}, 
\\*
\label{vndswi3}
z_s \,=& \,\,
\frac{(z- \bar z^{-1}) (\rho - C) + (z + \bar z^{-1}) 
(\rho_s - z \partial_z C - \bar z \partial_{\bar z} C)}
{2(\rho - C ) + (1+z \bar z)(\bar z \partial_{\bar z} C - \bar z^{-1} \partial_z C)}. 
\end{align}
\end{subequations}

Since (\ref{mrhieu1}) is the flat spacetime, we can write it with Cartesian coordinates as
\begin{eqnarray}
\label{mrhieu2}
\textrm{(\ref{mrhieu1})}=-dt^2 + dx^2 +dy^2 + dz^2.
\end{eqnarray}
We then formally write the supertranslated (\ref{mrhieu2}) as
\begin{eqnarray}
\label{mrhieu3}
ds^2=-dt^2 + dx_s^2 +dy_s^2 + dz_s^2,
\end{eqnarray}
where $t_s=t$ as in (\ref{vndswi}). 
(\ref{mrhieu3}) should agree with (\ref{metbwi}), 
therefore $dx_s^2 +dy_s^2 + dz_s^2$ should agree with $d\rho_s^2 + \rho_s^2\gamma_{AB} dz_s^Adz_s^B$, 
where $x_s^2+y_s^2+z_s^2=\rho_s^2$.
We can check these as
\begin{subequations}\label{hweor}
\begin{align}
\label{hweor1}
x_s^2+y_s^2+z_s^2 \,=& \,\, \rho_s^2, \\*
\label{hweor2}
dx_s^2+dy_s^2+dz_s^2 \,=& \,\,d\rho_s^2 + \rho_s^2\gamma_{AB} dz_s^Adz_s^B ,
\end{align}
\end{subequations}
with
\begin{subequations}\label{ctkiwqfe}
\begin{align}
x_s &= (\rho-C)\sin\theta \cos\varphi+\sin \varphi \csc \theta \,\partial_\varphi C-\cos\theta\cos\varphi \, \partial_\theta C, \\*
y_s &= (\rho-C)\sin\theta \sin\varphi- \cos \varphi \csc \theta \,\partial_\varphi C-\cos\theta\sin\varphi \, \partial_\theta C, \\*
z_s &= (\rho-C)\cos\theta+\cos\theta\cos\varphi \, \partial_\theta C, 
\end{align} 
\end{subequations}

In conclusion, if a 3D flat space is given, 
we can obtain that in the supertranslated coordinates 
by making the following replacement:
\begin{eqnarray}\label{veic}
(x, y, z) \to (x_s, y_s, z_s).
\end{eqnarray}

%============================================================ 
\section{Comment on the part of ``$\cdots$'' in (\ref{schcdst})}
\label{g34q} 
%============================================================

To be precise, (\ref{schcdst}) is denoted as
\begin{eqnarray}\label{bvuiwe}
ds^2
=
-( 1-{2 \mu(\rho)}/{r} ) dt^2 
+(1-{2 \mu(\rho)}/{r})^{-1}dr^2
+ 
j_{\theta\theta} d\theta^2
+ j_{\varphi\varphi} d\varphi^2
+
2j_{r\theta} dr d\theta,
\end{eqnarray}
where the term $2j_{r\theta} dr d\theta$ will appear through (\ref{trrlcx1}), 
and we may identify (\ref{bvuiwe}) with (\ref{rsso}). 
 (\ref{bvuiwe}) can be rewritten in  (\ref{schcdst}) with (\ref{ctfits}) as follows: 
\begin{eqnarray}\label{rvbei}
\textrm{(\ref{bvuiwe})}
\!\! &=& \!\!
g_{tt}dt^2
+g_{\rho\rho}d\rho^2
+\Big( 
j_{\theta\theta}+\frac{1}{1-\frac{2\mu}{r}}
\Big(
\frac{\partial r}{\partial \theta}\Big)^2
\Big)d\theta^2
+
\underbrace{
\frac{2}{1-\frac{2\mu}{r}}\frac{\partial r}{\partial \rho}\frac{\partial r}{\partial \theta}d\rho d\theta
+ 2j_{r\theta} dr d\theta}_{=0}
\nonumber\\*
&& \!\! +\, g_{\varphi\varphi}d\varphi^2,
\end{eqnarray}
where $dr$ is given as $\frac{\partial r}{\partial \rho}d\rho+\frac{\partial r}{\partial \theta}d\theta$, 
and $(\frac{\partial r}{\partial \theta}d\theta)^2$ is at the $\varepsilon ^2$-order. 
We may identify (\ref{rvbei}) with  (\ref{schcdst}). 
Note that cross terms can  exist in the expression with the $\rho$-coordinate, 
but canceled each other out in the expression with the $r$-coordinate obtained using (\ref{trrlcx1}).

%=============================================
\section{Derivation of (\ref{jiort}) from (\ref{sbdva})} 
\label{enlpu}
%============================================= 

We calculate the integral part in (\ref{sbdva}). 
\begin{align}\label{edvrn}
&\int_\Omega d^3\vec{x'} \frac{1}{|\vec{x}-\vec{x'}|}\frac{\cos \phi'}{r'\sin \theta'}T_{t\phi}(t,r',\theta')
\nonumber\\*
=&\,\, \int_\Omega d^3x' 
\frac{1}{r}(
1
+\frac{\vec{x} \cdot \vec{x'}}{r^2}
+\frac{1}{2r^2}(\frac{3(\vec{x} \cdot \vec{x'})^2}{r^2}-r'^2)
+{\cal O}(r^{-3})
)\frac{\cos \phi'}{r'\sin \theta'}T_{t\phi}(\vec{x'})
\nonumber\\*
=&\,\,
\frac{1}{r^2}\int_\Omega d^3\vec{x'} (\frac{\vec{x} \cdot\vec{x'}}{rr'}+\frac{3(\vec{x} \cdot \vec{x'})^2}{2r^3r'})\frac{\cos \phi'}{\sin \theta'}T_{t\phi}(\vec{x'})
+{\cal O}(r^{-4})
\nonumber\\*
=&\,\,
\frac{1}{r^2}\int_\Omega d^3\vec{x'}
\Big\{
\sin \theta \sin \theta'(\cos \phi \cos \phi'+\sin \phi \sin \phi')+\cos \theta \cos \theta' \nonumber\\*
&\qquad\qquad\quad +\frac{3r'}{2r}(\cos \theta\cos \theta'+\sin\theta\sin \theta'\cos(\phi-\phi') )^2
\Big\}
\frac{\cos \phi'}{\sin \theta'}T_{t\phi}(\vec{x'})
+{\cal O}(r^{-4})
\nonumber\\*
=&\,\,
\frac{1}{r^2}\int_\Omega d^3\vec{x'}
(\sin \theta \cos \phi \cdot \cos^2 \phi' 
+\frac{3r'}{r}\cos \theta\cos \theta'\sin\theta\cos \phi'\cos(\phi-\phi'))
T_{t\phi}(\vec{x'})+{\cal O}(r^{-4})
\nonumber\\*
=&\,\,
\frac{\sin \theta \cos \phi }{r^2}
\int dr' d\theta' 
\,r'{}^2 \sin \theta' 
\, T_{t\phi}(\vec{x'}) \cdot
\int_0^{2\pi} d\phi'' \cdot
\frac{\int_0^{2\pi} d\phi' \cos^2 \phi'} {\int_0^{2\pi} d\phi''}
\nonumber\\*
&
+\frac{3\pi \cos \theta \sin\theta \cos \phi}{r^3}
\int dr' d\theta' 
\,r'{}^3 \sin \theta'\cos \theta' \, T_{t\phi}(x') 
+{\cal O}(r^{-4})
\nonumber\\*
=&\,\,
\frac{\sin \theta \cos \phi}{2r^2}
\int_\Omega d\vec{x'}^3 
T_{t\phi}(\vec{x'})
+\frac{3\cos \theta \sin\theta \cos \phi}{2r^3}
\int_\Omega d\vec{x'}^3 \, r' \cos \theta' \, T_{t\phi}(\vec{x'}) 
+{\cal O}(r^{-4}), 
\end{align}
where 
$\vec{x}=(x,y,z)=r(\sin \theta \cos \phi, \sin \theta \sin \phi, \cos \theta)$ and $r=|\vec{x}|$. 
$\vec{x'}$ and $r'$ are likewise. 
We have exploited the fact that  $T_{t\phi}$ is independent of $\phi$ 
as the result of our assumption that the spacetime is axisymmetric toward $\phi$, 
which leads to the fact that $\int_0^{2\pi} d\phi'  \cos \phi'  \, T_{t\phi}(x')$ vanishes.  
From the result above, we can obtain (\ref{jiort}) from (\ref{sbdva}). 

%=============================================
\section{Detailed discussion of subtle points regarding the frame-dragging technique used in this study} 
\label{nowyt}
%============================================= 

In this appendix, we comment on the problem mentioned in Sec.\ref{nveoa} in detail. 
We have already mentioned the problem in Sec.\ref{nveoa} 
and the fact that the points where $\varphi$ and $\varphi_s$ appear are problematic.

Considering this fact, if we look at the equations in this work, 
we can see that the points where $\varphi$ and $\varphi_s$ appear are 
from (\ref{hoge}) to (\ref{bwenl}) and from (\ref{ldsrtr}) to (\ref{sklbwv}). 
Therefore, let us look at the analysis in those, respectively.  
\newline

Firstly, let us look at the analysis of (\ref{hoge}) to (\ref{bwenl}). 
The section ``$d\rho^2 + \rho^2 (d\theta^2 + \sin^2 \theta d\varphi^2)$'' in (\ref{hoge}) 
(referred to as the ``3D part'' from this point) can be regarded as 
a 3D flat space with $\rho$ as the radial coordinate, 
as in (\ref{bdml}). 
Therefore, as long as we focus on the 3D part and perform an analysis limited regarding that 3D part 
(for the intention of  ``limited regarding'', refer to the footnote in Sec.\ref{nveoa}), 
there is no problems with our considering $\varphi$ as an angular direction of the polar coordinate system in that part, 
regardless of whether it can be consistently obtained or not in the original 4D spacetime sense. 
Subsequently, if there are no problems in considering $\varphi$, 
there should be no problems either in considering the Cartesian coordinates $(x,y,z)$ 
instead of the polar coordinates $(\rho,\theta,\varphi)$ for that 3D part.

Then, (\ref{ctki0}) is given with $(x,y,z)$ and $(\rho,\theta,\varphi)$. 
Surely, $\varphi$ cannot be defined in the original 4D spacetime sense; 
however, as long as we use (\ref{ctki0}) only in the 3D part, 
there is no problems for the reason mentioned above. 
Actually, as can be seen in Appendix \ref{cwenclwe}, (\ref{ctki0}) is obtainable 
when a flat 3D space is generally given. 
We then explain that the following analysis from (\ref{ctki0}) is limited regarding the 3D part, in what follows. 
 
As can be seen in Appendix \ref{cwenclwe}, 
(\ref{ctki0}) is applicable to the 3D part,
and the analyses of (\ref{ctki0}) to (\ref{bwenl}) are always just substitutions and computations limited regarding the 3D part. 
Therefore, there is no problems in the analysis of (\ref{ctki0}) to (\ref{bwenl}).

Next, the 3D part in (\ref{yjtjr}) has been obtained only with $d\varphi$, as can be seen from (\ref{bwenl}); 
conversely, the 3D part in (\ref{yjtjr}) has been without $\varphi$ in (\ref{bwenl}). 
Therefore, problems with using  $\varphi$ do not exist in (\ref{yjtjr}), 
and it is possible to interpret (\ref{yjtjr}) in the original 4D spacetime sense. 
Then, if the 3D part in (\ref{hoge}) can be understood in the original 4D spacetime sense, 
the 3D part in (\ref{yjtjr}) can be understood in the original 4D spacetime sense as well, 
as the supertranslated (\ref{hoge}). 
Therefore, there is no problems in our using $\varphi$ in the analysis of (\ref{hoge}) to (\ref{bwenl}). 
\newline

Secondly, let us look at (\ref{ldsrtr}) to (\ref{sklbwv}). 
$\varphi_s$ appears in (\ref{ldsrtr}). 
However, that $\varphi_s$ is defined by the supertranslated coordinates $z_s$ and $\bar{z}_s$, 
which are generally defined in the 3D part as can be seen in (\ref{vndswi});
therefore, there should be no problems in our defining $\varphi_s$ in (\ref{ldsrtr}) 
as long as the following analysis from (\ref{ldsrtr}) is limited regarding the 3D part.

The computations of (\ref{ldsrtr}) to (\ref{lhrrvb3}) are always just substitutions 
toward $\theta_s$ and $\varphi_s$, as defined in (\ref{ldsrtr}). 
Therefore, there is no problems from (\ref{ldsrtr}) to (\ref{lhrrvb3}), 
if only $\varphi_s$ can be defined in (\ref{ldsrtr}).

Let us look at the rest (\ref{sklbwv}). 
``$\varphi \to \varphi_s$'' in (\ref{sklbwv}) works only as ``$d\varphi \to d\varphi_s$'' in effect, 
since there is no $\varphi$ in (\ref{aweq}). 
In addition, 
since we can obtain the relation between $\varphi$ and $\varphi_s$ as in (\ref{lhrrvb2}), 
which is $d\varphi=d\varphi_s$ in the approximation range of our analysis, 
the replacement (\ref{sklbwv}) has no impact on $d\varphi$ in (\ref{aweq}), in effect. 
Therefore, there is no problems with (\ref{sklbwv}), either. 
Subsequently, there is no problems in the analysis of (\ref{ldsrtr}) to (\ref{sklbwv}) as a whole. 
\newline

In conclusion, 
we have only used $\varphi$ and $\varphi_s$ in the parts that 
we can take out as independent parts 
(which means these parts are the discussion limited regarding the 3D part) from the complete analysis in this work. 
Moreover, we have never used them in the original 4D spacetime sense, at any points in this work. 
Therefore, there is no problems with our use of $\varphi$ and $\varphi_s$, in this analysis. 

Lastly, let us mention that the $C$ we have taken, 
which is the quantity for characterizing the type of the supertranslation to be involved, 
is independent of $\varphi$, as can be seen in (\ref{cdsvds}). 
Therefore, there is no chance for $\varphi$ to be involved in the analysis in this work.

%=============================================
\section{Expansion of (\ref{klafh}) at the infinite $r$} 
\label{rntre}
%============================================= 

We note the results of expansions of (\ref{klafh}) at the infinite $r$ as follows.
\begin{subequations}
\begin{align}
K_{tt} &= -1+\frac{2 m}{r}+{\cal O}(r^{-3}) +{\cal O}(r^{-4})\varepsilon+{\cal O}(\varepsilon^2),\\*
K_{t\phi} &= -\frac{2 a m \sin ^2\theta }{r}+{\cal O}(r^{-3}) +{\cal O}(r^{-2})\varepsilon+{\cal O}(\varepsilon^2), \\*
K_{rr} &= 1+\frac{2 m}{r}+{\cal O}(r^{-2})+{\cal O}(r^{-3})\varepsilon+{\cal O}(\varepsilon^2), \\*
K_{r\theta} &= ( -\frac{3}{4} \sqrt{\frac{5}{\pi }} m \sin (2 \theta )+{\cal O}(r^{-1}))\varepsilon +{\cal O}(\varepsilon^2),\\*
K_{\theta\theta} &= r^2+{\cal O}(r^{0})+ (3 \sqrt{\frac{5}{\pi }} m r \cos (2 \theta )+{\cal O}(r^{0}))\varepsilon +{\cal O}(\varepsilon^2),\\*
K_{\phi\phi} &= r^2 \sin ^2\theta +{\cal O}(r^{0})+ (\frac{3}{4} \sqrt{\frac{5}{\pi }} m r \sin ^2(2 \theta )+{\cal O}(r^{0}))\varepsilon+{\cal O}(\varepsilon^2).
\end{align}
\end{subequations}

%=============================================

\end{document}